\documentclass[aps,pra,superscriptaddress,reprint]{revtex4-1}

\usepackage{amsmath}
\usepackage{amssymb}
\usepackage{graphicx}
\usepackage{amsfonts}
\usepackage{bm}
\usepackage{mathrsfs}
\usepackage[mathscr]{eucal}
\usepackage{esint}
\usepackage{subfigure}
\usepackage{lineno}
\usepackage{esint}
\usepackage{subfigure}
\usepackage{supertabular}
\usepackage{booktabs}
\usepackage{amssymb}
\usepackage{latexsym}
\usepackage{CJK}
\usepackage{times}



\newcommand{\ket}[1]{\mid#1\,\rangle}
\newcommand{\bra}[1]{\langle\,#1\mid}
\newcommand{\ie}{{\it i.e.}}
\newcommand{\etc}{{\it etc.}}

\begin{document}

\title{Quantum Intelligence on Protein Folding Pathways}


\author{Wen-Wen Mao}
\affiliation{\noindent
Zhejiang Province Key Laboratory of Quantum Technology $\&$ Device and Department of Physics, Zhejiang University, Hangzhou 310027, P.R. China.}
\author{Li-Hua Lu}\email[email: ]{lhlu@zju.edu.cn}
\affiliation{\noindent
Zhejiang Province Key Laboratory of Quantum Technology $\&$ Device and Department of Physics, Zhejiang University, Hangzhou 310027, P.R. China.}
\author{Yong-Yun Ji}
\affiliation{\noindent
Department of Physics, Wenzhou University, Wenzhou 325035, P.R. China.}
\author{You-Quan Li}\email[email: ]{yqli@zju.edu.cn}
\affiliation{\noindent
Zhejiang Province Key Laboratory of Quantum Technology $\&$ Device and Department of Physics, Zhejiang University, Hangzhou 310027, P.R. China.}
\affiliation{\noindent
Collaborative Innovation Center of Advanced Microstructure, Nanjing University, Nanjing 210008, R.R. China.}

\keywords{protein folding $|$ quantum random walk $|$ shortest pathways $|$ mean-first passage time}

\received{\today}

\begin{abstract}
We study the protein folding problem on the base of the quantum approach we proposed recently
by considering the model of protein chain with nine amino-acid residues.
We introduced the concept of distance space and its projections on a $XY$-plane,
and two characteristic quantities, one is called compactness of protein structure and
another is called probability ratio involving shortest path.
Our results not only confirmed the fast quantum folding time
but also unveiled the existence of quantum intelligence hidden behind in choosing protein folding pathways.
\end{abstract}




\maketitle


Protein folding problem has been an important topic
in interdisciplinary field involving
molecular biology, computer science, polymer physics as well as theoretical physics etc..
Levinthal noted early in 1967 that a much larger folding time is inevitable
if proteins are folded by sequentially sampling of all possible conformations.
It was widely assumed
that a random conformational search does not occur in the folding process,
for which various hypotheses
with the help of a series of meta-stable intermediate states
have been often proposed.
There have been substantial theoretical models which are useful for understanding the essentials of the complex self-assembly reaction of protein folding with different
simplifying assumptions,
such as Ising-like model~\cite{Ising1,Ising2}, foldon-dependent protein folding model~\cite{Englander}, diffusion-collision model~\cite{Karplus,Sali}, and
nucleation-condensation mechanism~\cite{Thirumalai,Fersht}.
However till now, this often brings in certain difficulties in connecting analytical theory to experimental results
because some hypotheses can not be easily put into a practical  experimental measurement since they often  rely on various hypotheses\cite{Wolynes2005,Shakhnovich2006,Wolynes2012,Dill2012,Thirumalai2013}.
As the approach of computing simulations introduced less hypotheses
in comparison to those theoretical models, the atomistic simulations~\cite{Piana S,Henry R,Snow} have been also used to investigate the protein folding along with nowadays' advances in computer science.
All-atom computational method~\cite{Shahnovich2001}, including physics-based and knowledge-based approaches, have provided useful insights on protein folding and design by building high-accuracy atomistic models of proteins.
However, all those models are computationally costly for high-throughput folding studies and still need certain artificial hypotheses.

Recently, we introduced a quantum strategy~\cite{LuLi} to formulate protein folding as a quantum walk on a definite graph,
which provides us a general framework without making hypotheses,
where we merely studied the model with six amino-acids as toy model.
We know the shortest peptide chain in nature contains more than twenty amino-acid residues.
Toward a genuine understanding, it is obligatory to study  more complicate case with more residues.
Here we consider the model with nine amino-acid residues and obtain that
the folding time via our quantum approach is much shorter than the one obtained via classical random walk.
As the number of amino acid residues increases, the protein structure set becomes more complicate,
whereas, this drives us to introduce the projection method and find some new features on the protein folding pathways
in addition to the fast protein folding time.

\section*{Modeling}

\begin{figure}
\centering
\includegraphics[width=.8\linewidth]{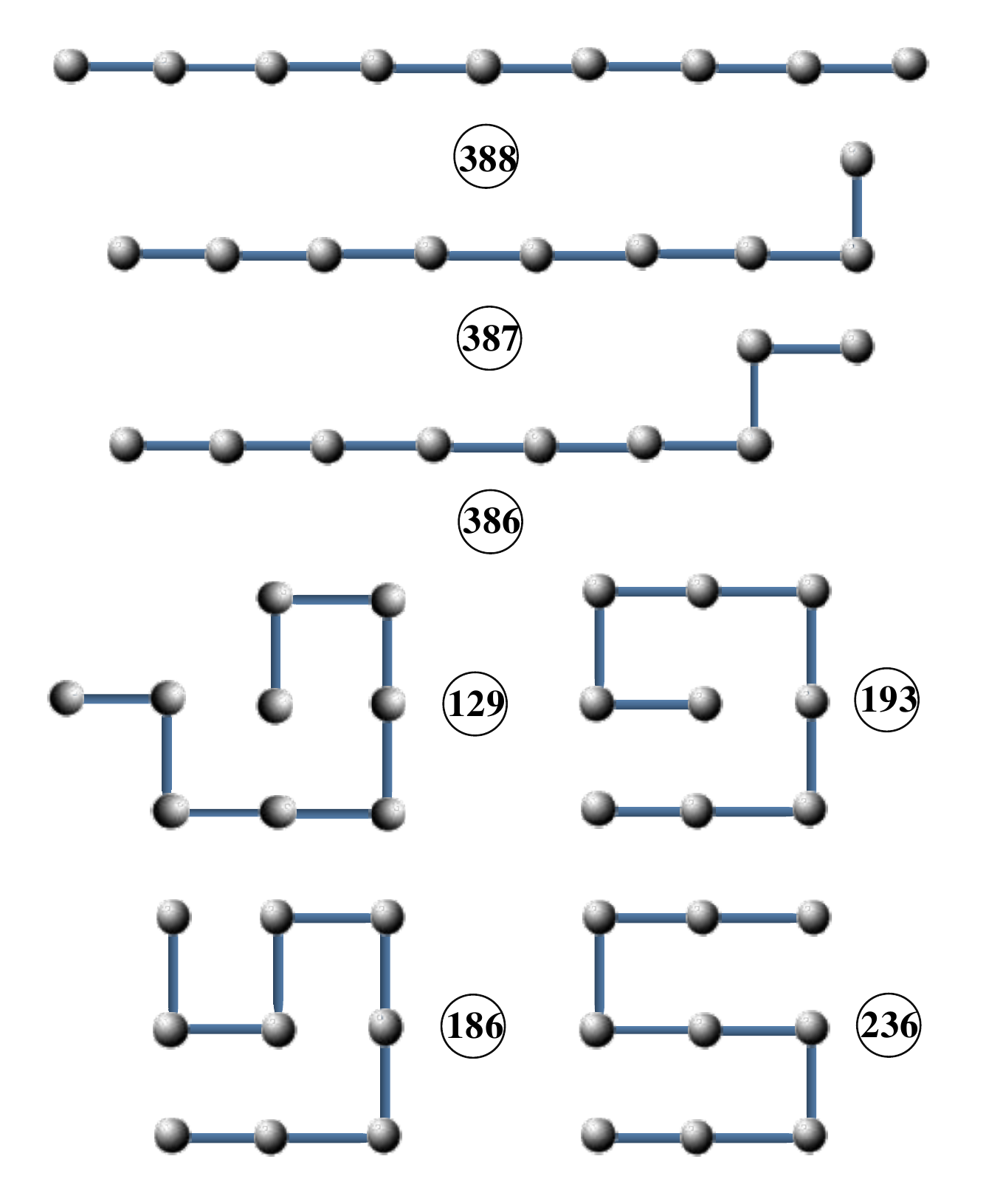}
\caption{
{\bf Structure Examples}
There are 388 distinct structures for the amino-acid chain with 9 residues, thus the corresponding structure set $\mathscr{S}_9$ contains 388 objects.
Here we plot the straight-line structure and the three most compact structures,
together with two examples related to them via a one-step folding.
For $n=9$ the connection graph $\mathscr{G}_9$ includes 388 sites which well defines the kinetic term of a quantum Hamiltonian.}
\label{fig:structure}
\end{figure}

In the course-grained model the protein is considered as a chain of non-own intersecting unit~\cite{GoModel,HP:Dill,LiTang1996,LiJi2004},
usually referring an amino acid residue of a given length on the two-dimensional square lattice.
We indicated recently~\cite{LuLi} that
for a protein with $n$ amino-acid residues, there will be totally $N_n$
distinct lattice conformations that distinguish various protein intermediate structures,
which provide us a point set with $N_n$ objects.
We call such a point set as structure set
and denote it by $\mathscr{S}_n =\{s^{}_1, s^{}_2, \cdots, s^{}_{N_n} \}$.
The entire structure set includes various structures that may be an unfolded, partially folded or completely folded.
To distinguish their difference, we introduce a concept of compactness of a structure,
\begin{equation}\label{eq:compact}
C=\frac{1}{n}\sum^{n}_{k=1}\bigl[(\mathbf{r}_k - \bar{\mathbf{r}})^2\bigr]^\frac{1}{2}
\end{equation}
where the geometric center of the protein is located at $\bar{\mathbf{r}} =(\mathbf{r}_{1}+\mathbf{r}_{2}+\cdots +\mathbf{r}_{n})/n$,
and $\mathbf{r}_k$ refers to coordinates of the $k$-th residues.
Clearly, the compactness $C_a$ represents the average distance from each residues to its geometric center of a given structure $s_a$.
We consider in this paper the case of $n=9$ and obtain $N_9=388$ through DSA-Depth Search Algorithm.
The $388$ distinct structures that are unrelated by rotational, reflection or reverse-labeling symmetries~\cite{LiTang1996}
were plotted in Appendix I.
The straight-line structure is labeled lastly as  $s_{388}$ for convenience
since it is the farthest end of the depth-first algorithm search.
We plot it together with the three most compact structures $s_{186}$,$s_{193}$ and $s_{236}$ in
Fig.~\ref{fig:structure} as an illustration.
For the structure set $\mathscr{S}_9$, the compactness of each structure is calculated and shown in the Appendix Table 2.
The largest value $C=2.222$ refers to the completely unfolded straight-line structure $s_{388}$, while the smallest value $C=1.073$  refers to the most compact structure $s_{186}$, $s_{193}$ and $s_{236}$.

On the basis of the lattice model,
we introduced previously
the concept of one-step folding to describe the protein folding process~\cite{LuLi}.
The one-step folding is defined by one displacement of an amino acid in one of the lattice sites: two protein structures are connected via one-step folding if their chain conformations differ in one site only.
For example, in Fig.~\ref{fig:structure}, the structures $s_{388}$ and $s_{386}$ can be obtained by a one-step folding from $s_{387}$,
and so did  $s_{193}$ from $s_{129}$.
This makes us to establish certain connections between distinct objects in the structure set $\mathscr{S}_9$
so that we have a connection graph $\mathscr{G}_9$ which is described by a $388\times 388$
adjacency matrix $\mathrm{Mat}(J_{ab})$ that characterizes a classical random walk~\cite{Kampen1997} on the connection graph.

\subsection{Distance space and its projections}

If the aforementioned graph $\mathscr{G}_n$ is completely connected, a distance between any two structures, saying $s^{}_a$ and $s^{}_b$, is well defined.
Then we will have a distance space
$\mathscr{D}_n:=\{ s^{}_a, d^{}_{a b}\}$
in which the magnitude of $d^{}_{a b}$ equals to the number of steps of the shortest path connecting $s^{}_a$ and $s^{}_b$ in the graph $\mathscr{G}_n$.
There has been four line-crossings already when the 22-vertex  graph $\mathscr{G}_6$ is plotted on a plane~\cite{LuLi}.
The $\mathscr{G}_9$ contains 388 vertices and appears very complicate if it is plotted on a plane.
The largest distance in $\mathscr{D}_9$ stretches between the completely unfolded straight-line structure $s_{388}$ and
the most compact structures $s_{186}$ or $s_{193}$, namely,
$d_{388,186}=d_{388,193}=17$.
While the distance between the structure $s_{388}$ and the other most compact structure $s_{236}$
is no more the farthest, it is just $d_{388,236}=14$.

It will be very helpful to make a projection of the space $\mathscr{D}_9$ into a $XY$-plane.
If mapping the $s_{388}$ to the origin point $(0,0)$ and either $s_{186}$ or $s_{193}$ to the farthest point at $(17,0)$,
we will have a set of points in the $XY$-plane where each point corresponds to one or several
objects in the space $\mathscr{D}_9$.
Their concrete locations in the $XY$-plane are determined by the the distance away from $(0,0)$  and that away from $(17,0)$, respectively.
The former represents the distance away from the straight-line structure $s_{388}$
and the later represents that away from either $s_{186}$ or $s_{193}$ in the distance space $\mathscr{D}_9$.

\begin{figure}
\centering
\includegraphics[width=.8\linewidth]{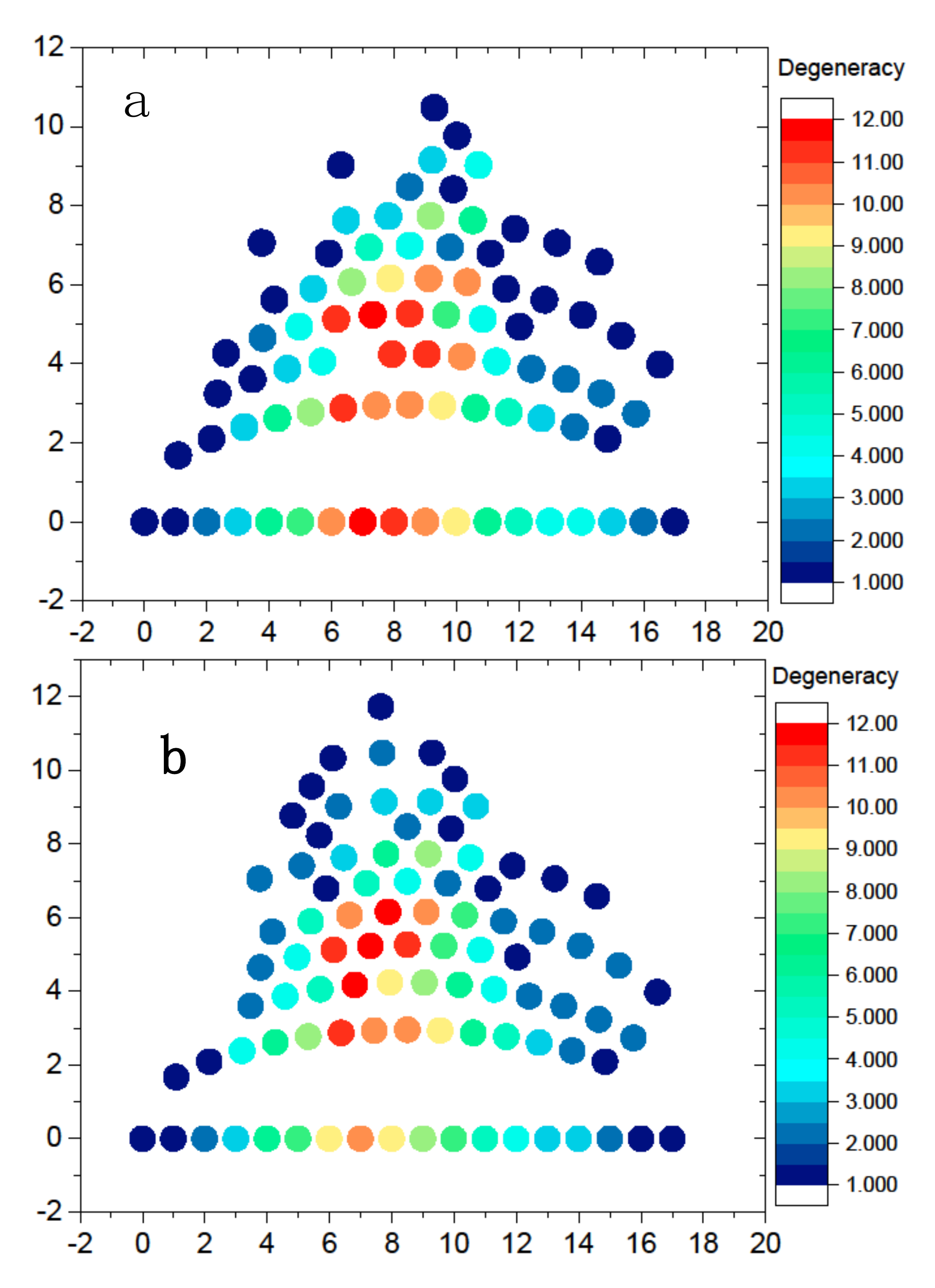}
\caption{{\small \bf Projection of the distance space} $\mathscr{D}_9$.
The color of the dot measures the degeneracy which is defined by the number of structures that concise on the same projection point.
The points lain on $X$-axis correspond to those structures on the shortest path.
Here $s_{388}$ was mapped to the origin point (0, 0), and
{\bf a,} $s_{186}$ as the farthest point,
{\bf b,} $s_{193}$ as the farthest point.}
\label{fig:projection}
\end{figure}

The color scatter diagram shown in Fig.~\ref{fig:projection} exhibits the aforementioned projection of $\mathscr{D}_9$ on $XY$-plane, in which
the color of each dot measures the degeneracy that is defined as the number of distinct structures mapping to the same point in the $XY$-plane.
The degeneracy of each dots in the above figure and the concrete structures contained in the original images
are all listed in Appendix {III}.
Those points lain on the $X$-axis in between $(0,0)$ and $(17,0)$ are the images of the objects referring to the shortest path leaving away from the initial structure $s^{}_{388}$ and approaching to a most compact structure.
Those objects on the shortest path together with the one mapped to $(17,0)$
constitute a subset $S_\mathrm{I}\subset\mathscr{S}_9$.
If let $S_\mathrm{\small I\!I}$ denote the set constituted by the other objects,
then the structure set $\mathscr{S}_9$ can be partitioned into two subsets $S_\mathrm{\small I}$ and $S_\mathrm{\small I\!I}$
namely, $\mathscr{S}_9= S_\mathrm{\small I}\oplus S_\mathrm{\small I\!I}$.

The density matrix $\rho(t)$
describing quantum dynamics is of $388\times388$ that is too large to manifest some useful information.
The above projection picture helps us to properly reduced the large density matrix to a $2\times 2$ one
to extract certain useful information, namely,
\begin{equation}
\tilde{\rho}=
\left(
   \begin{array}{cc}
   \rho^{}_\mathrm{\small I, I}  &  \rho^{}_\mathrm{\small I, I\!I} \\[1mm]
   \rho^{}_\mathrm{\small I\!I, I}  &  \rho^{}_\mathrm{\small I\!I, I\!I}
   \end{array}
\right)
\label{eq:Reducedmatrix}
\end{equation}
where
$\tilde{\rho}^{}_\mathrm{\small I, I} = \sum_{a}\rho^{}_{a a}$,
$\tilde{\rho}^{}_\mathrm{\small I\!I, I\!I} = \sum_{b}\rho^{}_{b b}$,
$\tilde{\rho}^{}_\mathrm{\small I, I\!I} = \sum_{a,b}\rho^{}_{a b}$
and
$\tilde{\rho}^{}_\mathrm{\small I\!I, I} = \tilde{\rho}^{*}_\mathrm{\small I, I\!I}$
with
$a\in S_\mathrm{\small I}$ and $b\in S_\mathrm{\small I\!I}$.
Here the large density matrix stands for $\rho(t)=\ket{\Psi(t)}\bra{\Psi(t)}$.

\subsection{On the time evolution}

Letting $\ket{s_a}$ denote the state of a protein structure in the shape of the $a$-th lattice conformation,
we will have a quantum Hamiltonian in a $388$-dimensional Hilbert space
$\mathscr{H}=\{ \ket{s^{}_a} \mid a=1, 2, \cdots, 388 \}$, namely,
$\hat{H} =-\sum_{a,b} J_{a b}\ket{s^{}_a}\bra{s^{}_b}$
where $J_{a b}$ refers to the connections between different objects in the structure set $\mathscr{S}_9$,
\ie, $J_{a b}$ is not zero if the $a$-th protein structure can be transited into the $b$-th one by a one-step folding while vanishes otherwise.
With these physics  picture one can also investigate quantum walk~\cite{Aharonov1993,Farhi1998,WangJB2014} on the aforementioned graph, which gives a  quantum mechanical understanding of the protein folding, \ie, the process by which proteins achieve their native structure.
As the time evolution is governed
by the Schr\"odinger equation
\begin{equation}
i\hbar\frac{\mathrm{d}}{\mathrm{d}t}\ket{\Psi(t)}=\hat{H}\ket{\Psi(t)}
\label{eq:time}
\end{equation}
in which
$\ket{\Psi(t)}=\sum^{388}_{a=1}\psi_a (t)\ket{s^{}_a}$,
and the expansion coefficients $\psi_a(t)$ can be solved numerically at least.
In our numerical calculation, we set $\hbar$ and $J$ to be unity
and take the time step as $\Delta t=0.02$.
For the initial condition,
$\ket{\Psi(0)}=\ket{s_{388}}$,
we solve the first-order differential equation [\ref{eq:time}]
by means of Runge-Kutta method and obtain the magnitude of $\psi_a (t)$ at any later time,
$t=j*\Delta t$  with $j=1,2,\cdots$.


In order to compare with the protein folding problem in the classical literature,
\ie, a random conformation search process,
let us revisit the classical random walk.
The continuous-time classical random walk~\cite{Farhi1998} on a graph
$\mathscr{G}_n$ is described by the time evolution of the probability distribution
$p^{}(t)$ that obeys the equation of motion
\begin{equation}\label{eq:classicalRW}
\frac{\mathrm{d} }
     {\mathrm{d}t}
      p^{}_a (t) = \sum_b K^{}_{a b}\,p^{}_b (t)
\end{equation}
where  $K_{a b}=T_{ab}-\delta_{a b}$
with $T^{}_{ab}$ being the probability-transition matrix.
In the conventional classical random walk,
the probability-transition matrix is determined by the adjacency matrix of an undirected graph,
namely,
$T_{ab}=J_{ab}/\mathrm{deg}(b)$
where $\mathrm{deg}(b)=\sum_c J_{cb}$ represents the degree of vertex-$b$
in the graph.

\section*{Results}

It is widely believed that the native structure of a protein possesses the lowest free energy~\cite{Anfinsen}.
This can be interpreted by the hydrophobic force that drives the protein to fold into a compact structure with as many hydrophobic residues inside as possible~\cite{HP:Dill}.
How fast does a protein initially in a straight-line structure folds into the most compact structure
and what is the main behavior during the folding process
will be important issues to study.

\subsection{Folding time and folding pathway}
We let $P^{}_{a,b}(t)$ denote the probability of a state being the basis state $\ket{b}$ at time $t$ if starting from the state $\ket{a}$ at initial time $t=0$.
Quantum mechanically, $P_{a,b}(t)= |\psi^{(a)}_b(t)|^2$ as long as we solved
the Schr\"odinger equation [\ref{eq:time}] in terms of
the initial condition $\ket{\Psi(0)}=\ket{s_a}$.
Here the superscripts are introduced to distinguish the solution from different initial conditions.
As we known,
the first-passage probability $F^{}_{a,b}(t)$ from a state $\ket{s^{}_a}$ to another state $\ket{s^{}_b}$ after $t$ time obeys the known convolution relation~\cite{FirstPassageTime1969,FirstPassageTime2001,FirstPassageTime2004,FirstPassageTime2007,FirstPassageTime2016}
\begin{equation}\label{eq:FPTprobability}
P^{}_{a,b}(t)=
\int^{t}_{0} F^{}_{a,b}(t')P^{}_{b,b}(t-t')\mathrm{d}t'
\end{equation}
where $P_{b,b}(t)= |\psi^{(b)}_b(t)|^2$ arises from the solution of Eq.~[\ref{eq:time}] from
another initial condition $\ket{\Psi(0)}=\ket{s_b}$.
In the classical case, $P^{}_{a,b}$ and $P^{}_{b,b}$ refer to the $p^{}_{b}(t)$ solved from equation [\ref{eq:classicalRW}], respectively, with initial conditions $p^{}_c (0) =\delta^{}_{ac}$
and $p^{}_c (0) =\delta^{}_{bc}$.

As protein folding is the process that proteins achieve their native structure,
the folding time is the case that the starting state is chosen as $\ket{s^{}_{388}}$ and the target states are the most compact states $\ket{s^{}_c}$.
For example, they are $\ket{s^{}_{186}}$, $\ket{s^{}_{193}}$
for $n=9$ as aforementioned.
With the help of the first-passage probability solved from [\ref{eq:FPTprobability}],
we can calculate the folding time by the formula
\begin{equation}
\tau^{}_\mathrm{fd}=\displaystyle\frac{
 \int_{0}^{\tau^{}_0 } t F^{}_{a,c}(t)\mathrm{d}t }{
      \int_{0}^{\tau^{}_0 }  F^{}_{a,c}(t)\mathrm{d}t}
\end{equation}
where $\tau^{}_0$ represents the time period when the first-passage probability vanishes $F_{a,b}(\tau^{}_0)=0$~\cite{LuLi}.
Here $a=388$ and $c=186$ or $193$.
We know in a previous work that the quantum folding is faster than the classical folding with about four to six times even for the simplest model of 4 residues and it is faster than the classical folding with almost ten to hundred times or more for $n=6$~\cite{LuLi}.
Here we calculate the folding time for $n=9$  that is given in Table 1.
We can see that the quantum folding time $\tau^\mathrm{q}_\mathrm{fd}$
is much shorter than the classical folding time $\tau^\mathrm{c}_\mathrm{fd}$,
and their difference becomes more significant in the of $n=9$ in comparison to $n=6$.

\begin{figure}[t]
\includegraphics[width=0.5\textwidth=1]{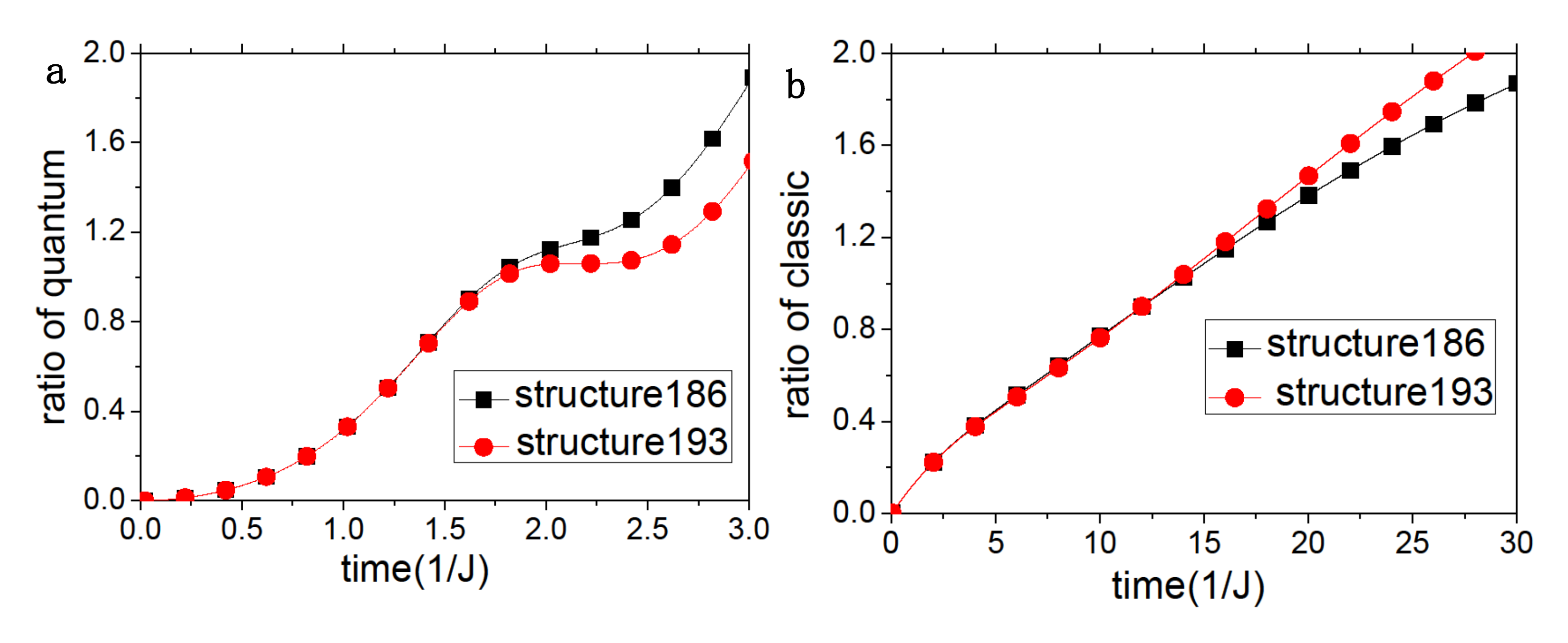}
  \vspace{-2mm}
\caption{\label{fig:ratiograph}\small
{\small \bf The probability ratio $\gamma$.}
The time evolution of $\gamma$ in the folding process from $s_{388}$ to $s_{186}$ and $s_{193}$ for
{\bf a,} quantum walk and for {\bf b,} classical random walk.}
\end{figure}

In order to explore whether there are any intelligence hidden behind
in choosing the protein-folding pathway,
we compare the total probability on the shortest path and the other path by
a probability ratio $\gamma$ to demonstrate how they changes relatively
after it leaves the initial state, namely,
\begin{equation}
\gamma(t)=\frac{\Gamma_\mathrm{\small I\!I}(t)-P^{}_{a,a}(t)}{\Gamma_\mathrm{\small I}(t)}
\end{equation}
where
$\Gamma_\sigma(t)=\sum_{b\in S_\sigma}P^{}_{a,b}(t)$
with
$\sigma=\mathrm{\small I}, \mathrm{\small I\!I}$.
Our result is plotted in Fig.~\ref{fig:ratiograph}, we can see that
the magnitude of $\gamma$ changes from $0$ to $1$ monotonously during a short period of time
for both quantum and classical cases.
This implies that the probability on the shortest path is more favorable after leaving the initial state
for it is on the denominator of $\gamma$.
In order to rule out the ambiguity on the time scales about classical and quantum literature,
we evaluate a mean ratio,
$\bar{\gamma}=  \int_{0}^{\gamma=1 }\gamma(t)\mathrm{d}t/
      \int_{0}^{\gamma=1 }\mathrm{d}t$
and show them in Table~\ref{table1}.
\begin{table}[h]
  \centering
  \caption{The probability ratio and the folding time}
  \label{table1}
    \begin{tabular}{crrrrr}
    \toprule
   $\mathrm{structure}$ & $\bar{\gamma}^\mathrm{q}$ & $\bar{\gamma}^\mathrm{c}$ & $\tau^\mathrm{q}_\mathrm{fd}$ & $\tau^\mathrm{c}_\mathrm{fd}$ \\[1mm]
    \midrule
    {$s^{}_{186}$}  & 0.341002 & 0.551697 & 8.549224 & 2153.938 \\[1mm]
    {$s^{}_{193}$}  & 0.353285 & 0.542997 & 6.440695 & 3305.09 \\[1mm]
    \bottomrule
    \end{tabular}%
  \label{tab:addlabel}%
\end{table}%
The mean ratio in quantum case $\bar{\gamma}^\mathrm{q}$ is smaller than that in classical case $\bar{\gamma}^\mathrm{c}$.
This implies that the probability distribution is more aggregated on the shortest path in quantum case.

\subsection{More on folding behavior}

Although the random walk reflects a stochastic process,
it is characterized by the probability distribution defined on all the distinct structures
that can always give us intuitive information if we calculate some weighted average of the entire system.
We can attain the time dependence of mean distance  away from any structure $s_a$ by calculating $\sum_b d^{}_{a b}|\psi^{(a)}_b(t)|^2 $.
We can also observe the time evolution of the mean compactness $\sum_b C^{}_a|\psi^{(a)}_b(t)|^2 $.
In classical case, the mean distance and mean compactness is evaluated by $\sum_b d^{}_{a b}p^{(a)}_b(t)$
and $\sum_b C^{}_ap^{(a)}_b(t)$ respectively.

\begin{figure}[t]
\includegraphics[width=0.5\textwidth=1]{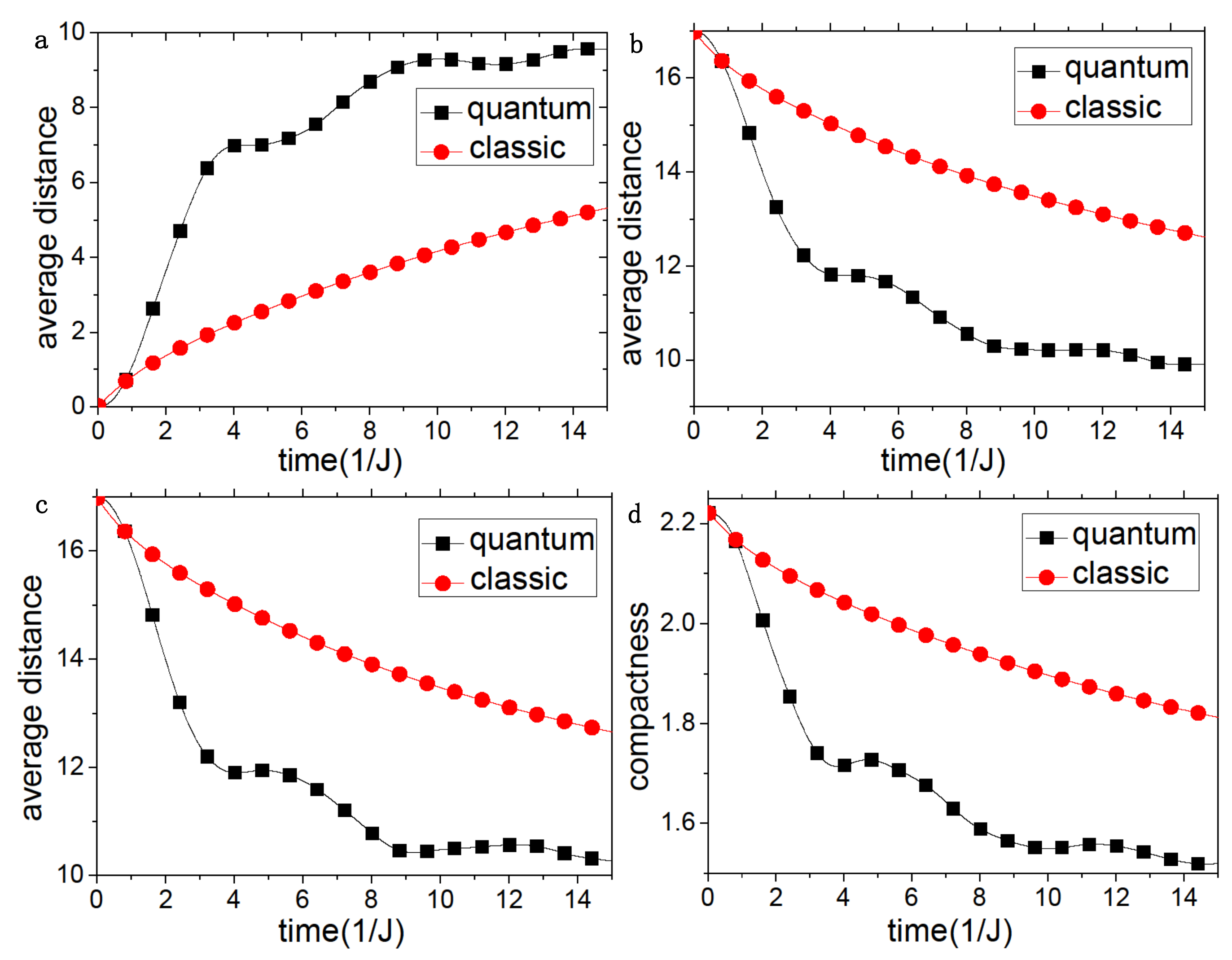}
\caption{\label{fig:average}\small
{\small \bf Time evolution of average distance and average compactness.}
{\bf a,} The average distance leaving away from the initial structure $s_{388}$.
{\bf b,} The average distance approaching to the most compact structure $s_{186}$
         while leaving from the structure $s_{388}$.
{\bf c,} The average distance approaching to the most compact structure $s_{193}$
         while leaving from the structure $s_{388}$.
{\bf d,} The time dependence of the overall average of compactness.}
\end{figure}

Our calculation of those two average quantities are given in Fig.~\ref{fig:average}.
We can see that the mean distance of leaving away from the initial straight-line structure $s_{388}$
increases much more rapidly in the quantum folding process than in classical case ( see in Fig.~\ref{fig:average}~{\bf a}).
Correspondingly,
the mean distance approaching to the most compact structures $s_{186}$ and $s_{193}$
decreases more rapidly (see in Fig.~\ref{fig:average}~{\bf b,c}) in the quantum folding process than in classical one.
Additionally,
the average of compactness also decreases more rapidly in the quantum folding process  than in class  case,
(see Fig.~\ref{fig:average}~{\bf d}),
thus the protein classically shrinks slower
and it quantum mechanically shrinks faster.

As we know, quantum mechanically, the off-diagonal elements of a density matrix reflects
certain quantum coherent properties, but our discussion till now have not yet involve it directly.
The partition of entire structure set into subset by considering the concept of shortest path
help us to have a $2\times 2$ density matrix.
It it then worthwhile to calculate von Neumann entropy~\cite{Nielsen2001}
$E_\mathrm{\small N}(\tilde{\rho})\,=\,-\mathrm{tr} (\tilde{\rho} \,{\rm {\log_2}}\tilde{\rho})$.
We also calculated the Shannon entropy~\cite{Pathria2011}
$E_\mathrm{\small S}
 =-p^{}_\mathrm{\small I}\log p^{}_\mathrm{\small I}
  -p^{}_\mathrm{\small I\!I}\log p^{}_\mathrm{\small I\!I}$
for the classical result to compare.
Here $p^{}_\mathrm{\small I}=\sum_{a\in S^{}_\mathrm{\small I}} p^{}_a$
and
$p^{}_\mathrm{\small I\!I}=\sum_{b\in S^{}_\mathrm{\small I\!I}} p^{}_b$.
The aforementioned  $2\times 2$ density matrix $\tilde{\rho}$ can also define a quantity called the degree of coherence~\cite{Leggett,LuLi2009}
$\eta(t)=2\mathrm{tr}\tilde{\rho}(t)^2-1 $,
which is related to various quantum coherence phenomena, such as Rabi osicillation, self-trapping \etc.
After making some calculus, one can found that the degree of coherence and the Von Neumann entropy reach the extreme value  at the same time,
precisely, $\eta(t)$ takes minimal value when $E_\mathrm{\small N}(t)$ takes its maximum.

\begin{figure}[t]
\includegraphics[width=0.5\textwidth=1.6]{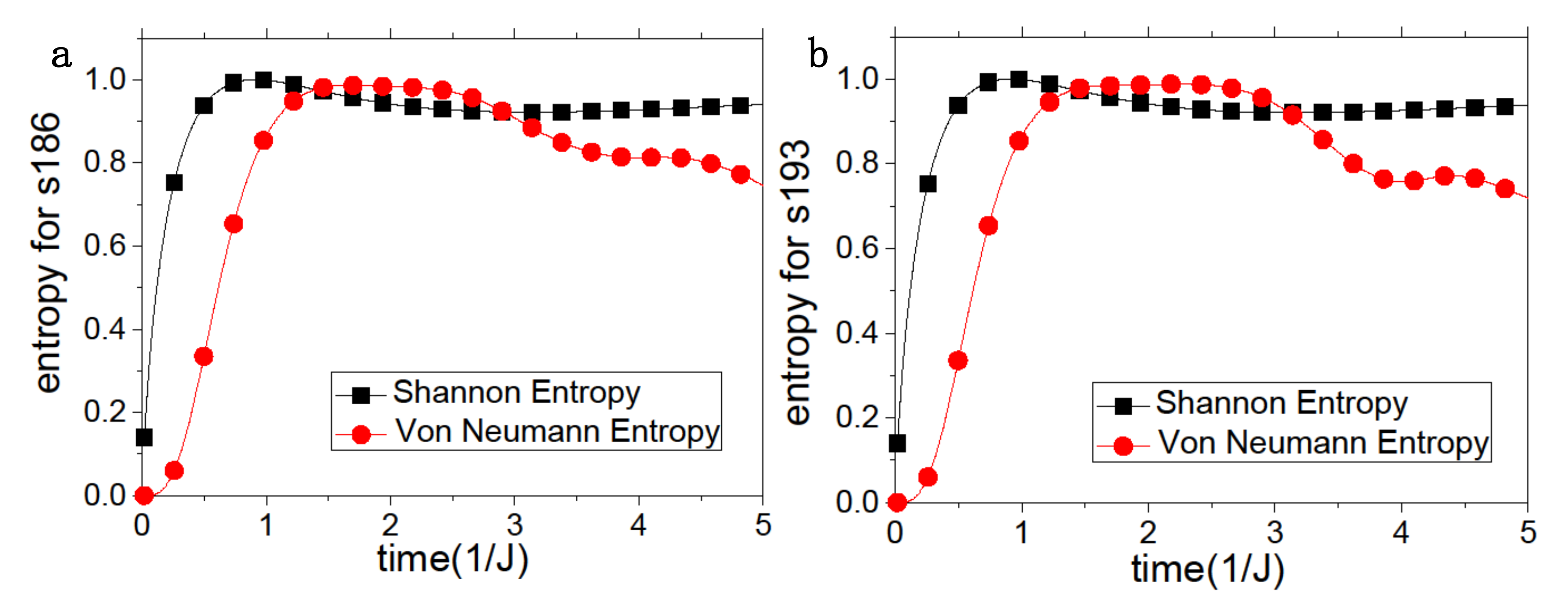}
  \vspace{-2mm}
\caption{\label{fig:entropy}\small
{\small \bf Time-dependent evolution of entropy }
{\bf a,} The quantum and classic time evolution of the entropy from $s_{388}$ to $s_{186}$,
{\bf b,} The quantum and classic time evolution of the entropy from $s_{388}$ to $s_{193}$.}
\end{figure}

The time dependence of both Von Neumann and Shannon entropies are plotted in Fig.~\ref{fig:entropy}.
We can see that both entropies change from $0$ monotonously to a maximal value and then varies.
The maximal Von Newmann entropy means the quantum state is in a completely mixed state (the degree of coherence vanishes),
while a maximal Shannon entropy implies the maximal information uncertainty.
In our present problem,
the population (or probability summation in classical case) on the shortest path is dominated during the time before the entropy reaches the maximal value.
Clearly, such a period in quantum case is longer than in classical case.
This illustrates from another angle of view that quantum folding process posses more intelligence in protein folding pathway.

\section*{Summary}

In the above, we studied the protein folding problem on the base of the quantum approach we proposed recently~\cite{LuLi}
by considering the model of protein chain with nine amino-acid residues.
We show that the protein folding can be modelled as a quantum walk on the graph $\mathscr{S}^{}_9$ of $388$ vertices.
As such a graph appears complicate if it were plotted on a plane, we introduced
the concept of distance space $\mathscr{D}_9$ and its projections on a $XY$-plane by choosing two farthest structures
(one is the completely unfolded straight-line structure and the other is a most compact structure) as
reference-base points.
According to our scheme~\cite{LuLi}, we obtained the protein folding time by making use of the first-passage probability.
In order to attain more understandings on the folding behavior,
we introduced two characteristic quantities, one is called compactness of protein structure
another is called probability ratio involving shortest path.
The introduction of the concept of shortest path help us to reduce the $388\times 388$ large density matrix to a $2$ by $2$
density matrix. Then we can conveniently evaluate the Von Neumann entropy etc..
We also calculated the Shannon entropy on the base of classical random walk approach to compare.
Our results not only confirmed the fast quantum folding time~\cite{LuLi}
but also unveiled the existence of a quantum intelligence hidden behind in the choosing protein folding pathways.


This work is supported by National Key R \& D Program of China, Grant No. 2017YFA0304304, and partially by the Fundamental Research Funds for the Central University.


\bibliography{50}

\end{document}